\documentclass[twocolumn, 10pt]{article}
\usepackage{amsmath, amsfonts, graphicx,epsfig}
\usepackage{subfigure,times}

\date{May, 2005}

\title{\Large\bf Reverse Engineering Point Clouds to Obtain Tensor Product 
B-Spline Surfaces by Blending Local Fits}

\author{
	Lavanya Sita Tekumalla\\
	\em{University of Utah}\\
	\em{lavanyat@cs.utah.edu}
	\and 
	Elaine Cohen \\
	\em{University of Utah}\\
	\em{cohen@cs.utah.edu}}
	
\begin{document}
\maketitle
\setlength{\topmargin}{0.0in}
{\normalsize

\thispagestyle{empty}

\begin{abstract}
\textit{Being able to reverse engineer from point cloud data to obtain 3D 
models is important in modeling. As our main contribtion, we present a new method 
to obtain a tensor 
product B-spline representation from point cloud data by fitting surfaces to 
appropriately segmented data. By blending multiple local fits our method is more 
efficient than existing techniques with the ability to deal with a detail by efficiently 
introducing a high number of knots.
Further point cloud data obtained by digitizing 3D 
data, typically presents many associated complications like noise and missing 
data. As our second contribution, we propose an end-to-end framework for 
smoothing, hole filling, 
parameterization, knot selection and B-spline fitting that addresses these issues, 
works robustly with large irregularly shaped data containing holes and is straightforward 
to implement.}
\end{abstract}
\section{Introduction}
Digitizing 3D data and reverse engineering data to obtain 3D tensor product 
spline models has numerous applications in the field of CAD modeling and 
entertainment. Hence, the problem of automating design for manufacture and 
production by converting real world objects into computer models is extremely 
important. This paper focuses on processing noisy incomplete point clouds obtained from 
scanners and converting them to a tensor product spline representation with parameter 
values representative of the objects features.

3D data from real world objects obtained by scanning models is normally 
associated with several problems like noise in the data, missing  data and 
holes. Incomplete data and holes in the data might typically lead to 
ill-behaved surfaces. Though surface in the hole regions is finally trimmed away after 
fitting, the area around holes might be affected and rank deficiencies might 
be encountered in the fitting phase. Further, most of the parameterization 
techniques do not deal with incomplete data.

\textbf{ Contributions: } We propose a novel and efficient B-Spline
surface fitting algorithm that blends local fits on segments of the data, 
with an automated knot placement strategy. We propose an end-to-end framework 
to reverse engineer point clouds
with 
noisy data and holes in the data 
and fit tensor product B-spline surfaces that involves smoothing, hole filling, 
parameterization, knot selection and B-spline surface fitting steps.

We assume that an underlying triangular mesh structure is available with the data
 to obtain a good parameterization of the data. 
Where such information is not available, one could use existing meshing algorithms. 
We consider address models with an arbitrary number of 
holes, with the assumption that there is reasonable sampling density where 
triangulated data is available. We deal with a single patch of a segmented 
model that is homomorphic to a disc, with an arbitrary number of holes. Thus 
we do not aim to deal with issues such as approximating sharp features in the 
input and assume that such features do not exist in the input.  

\section{Prior Work}
Though several aspects of the reverse engineering problem such as data 
parameterization, data fitting and knot selection have been extensively 
explored, there has not been significant work  dealing with an entire 
pipeline for such a process for ill-conditioned input and data with holes in 
the input. B-spline surface fitting involves the steps of  
parameterization, knot placement and the actual fitting algorithm. 

The task of parameterization involves finding a one to one mapping from every 
point on the surface to a point in a parametric domain. A lot of work has 
been done in this area to obtain planar maps for triangular meshes. A 
detailed discussion on these methods can be found in \cite{FH}. Once an 
initial mapping has been obtained, the parameterization can be improved using 
various criteria. In \cite{HL,PS,CO} iterative parameter correction is used 
to reparameterize the surface. In \cite{SGS}, the surface is  reparameterized 
based on a stretch metric that measures the distortion in scale.

The traditional criterion for approximating data when there is more data than 
the degrees of freedom is the linear least squares method that minimizes the 
$L^2$ norm of the residual. The weighted least squares technique is a related 
method that minimizes the least squares error and at the same time associates 
a weight with each error term based on its relative importance. One way of 
assigning weights while performing weighted least squares, is to define the 
weighting function with respect to a specific focus point . Therefore a 
different fit is obtained every time this point is changed. This method of 
weighting points is the basis for the MLS projection method described in 
section \ref{MLS-proj}.

Several methods have attempted to address the additional problem of noise 
while fitting scattered data. One way to deal with scattered data is to 
minimize a combination of the $L^2$ norm and the smoothing norm \cite{GH,DH}.

Knot placement has a significant effect on the quality of the resulting 
surface during fitting. Several attempts \cite{BR,DJ,DP} have been made to 
consider the least squares problem as a nonlinear optimization problem where 
the position of the knots is also optimized along with the control points. In 
\cite{BMP}, an iterative procedure is presented that inserts and deletes 
knots adaptively. However all these methods are computationally expensive and 
take significant execution time.

There exists some work on outlining an end-to-end framework for fitting B-spline 
surfaces to point clouds. In \cite{GH}, a method to approximate scattered 
data that is triangulated using a single tensor product B-spline patch or 
hierarchical B-splines is presented. They minimize a functional that is a 
combination of the least squares distance and a fairing term based on data 
dependent thin plate energy in the fitting step. In \cite{DH}, a framework is 
given for reverse engineering point clouds to get trimmed NURBS in which they 
too use a fairness term while minimizing error for fitting. However they 
parameterize data by projection and hence are not assured of finding a one to 
one mapping for all geometry.

Work has been done that attempts to deal with an entire model by making a 
network of B-spline patches. A method to construct B-spline models of 
arbitrary topology, by constructing a quadrilateral network of B-spline 
patches obtained by merging triangles in the input mesh appropriately is 
presented in \cite{EDD}.  In \cite{KL}, another procedure is presented, where 
the user interactively segments the data. In \cite{PYL}, a method is 
described in which the model is segmented  using K-means clustering and then 
a NURBS patch network is computed. \cite{GHJ} decomposes the given point set 
into a quad-tree like data structure called a strip tree, to construct the patch network.
Though some of these methods deal with the complex problem of handling the 
entire model, unlike this paper which deals with a single patch, none of 
these methods deal with holes and missing data. Further, our surface fitting technique 
of blending local fits to form a global fit is computationally efficient even with a large
number of knots. Hence our method can be used used to effectively deal with a higher level 
of surface detail, an important requirement in any realistic application.

\subsection{Background : MLS Projection Procedure}
\label{MLS-proj}
In this section, we briefly review the MLS projection that is used through the various 
steps in our pipeline.
The MLS projection procedure was proposed by \cite{L} and \cite{ABC} to deal 
with meshless surfaces. Given a point set, the MLS projection operator 
projects a point $r$ near the surface onto the surface implicitly defined by the set of points. This surface can be defined as the set of points that 
project onto themselves. 

The MLS projection operator proceeds in two steps. To project a point $r$, 
the first step requires finding an optimal local reference plane for the 
neighborhood of $r$ by minimizing the $L^2$ norm of the weighted perpendicular distance of points $p_{i}$ in the neighborhood from the optimal reference plane. If $n$ is the normal to the plane and $t$ the 
distance of the plane from $r$ (figure \ref{mls-surf}), $\sum\limits_{i = 1}^N 
{\left\langle {n,p_{i} - r - tn} \right\rangle } ^2 \theta (||p_{i} - r - 
tn||)$ is minimized with respect to $n$ and $t$, where $\theta$ is a gaussian 
weighting function defined as $\theta(x)=e^{(-x^2/h)}$. This is a non-linear 
minimization process. A local parameterization is obtained by projecting each 
point in the neighborhood onto this reference plane.  The next step involves 
fitting a local bi-quadratic polynomial surface $g$ using the moving least 
squares technique. That is, we find a $g$ to minimize $\sum\limits_{i = 1}^N 
{(g(x_{i}},y_{i}) - f_{i})^2 \theta (||p_{i} - q||)$ where $q = r + tn$ ($q$ 
is the projection on the best fit plane), $(x_{i},y_{i})$ are the parameter 
values of $p_{i}$ in the local reference plane and $f_{i}=<p_{i}-q,n>$ is 
orthogonal to the local reference plane.This polynomial when evaluated at the 
point $q$, gives the desired MLS projection. 

We use the MLS projection procedure for smoothing and for filling holes in 
our framework.

\begin{figure}[t!]
\centering
\epsfxsize=2.5in
\epsffile{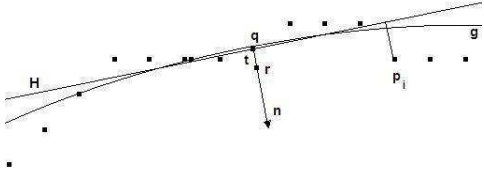} 
\caption{MLS Projection procedure for surfaces}
\label{mls-surf}
\end{figure}

\section{Our Reverse Engineering Technique}
We now present a framework that makes tensor product B-spline surfaces from triangulated point clouds. 
We propose a movel B-spline surface fitting algorithm based on blending local fits and  an automated 
knot selection strategy in a later part of the section. 
We compare our fitting 
method with the traditional global least squares fit in terms of quality of 
fit and computation speed in the experiments section. 
Also, the effects of certain parameters of our 
method on the quality of the resulting fit, such as the choice of 
neighborhood size and the weighting function, are discussed.

In order to deal with noisy and incomplete data, 
we take a multi-stage approach of smoothing, hole filling, 
parameterization, knot selection and fitting. Each of these steps is detailed 
in the following sections. 

\subsection{Smoothing}
The smoothing step removes the noise and outliers in the data. Smoothing is 
achieved by projecting each point in the point cloud onto the MLS surface 
computed at that point. However, this method smooths the data in the local 
normal direction obtained through the MLS method. As a result, the boundary 
curve may contain significant noise even after smoothing the surface. That 
is, normals of the points in the boundary curve may not lie along the surface 
normals at these points. Hence, normal smoothing of the surface alone does 
not suffice.

We introduce an additional step of smoothing the boundary by projecting each point 
in the boundary onto the MLS curve computed locally at the point \cite{TCSC}. 
Suppose $r$ is a noisy point near the curve, and $q$ is its projection onto the 
line with direction vector $u$, we find the optimal line such 
that $\sum\limits_{i = 1}^N {||(p_{i} - q) - \left\langle {p_{i} - q,u} \right\rangle u||} ^2 \theta (||p_{i} - q||)$ 
is minimized with respect to $q$ and $u$. Again, this is a non-linear minimization process. 
The data is parameterized locally by projecting it onto the line along $u$ and passing through $q$. 

In the next step, the point $p_{i}$ is projected onto a local quadratic 
approximation of the curve, in a process similar to \cite{ABC}. First a local 
coordinate system is found, with $u$, $(r-q)$ and $u \times (r-q)$ as the 
axes. Then the local neighborhood is transformed to this system. In 
order to handle 3D data, we treat the local curve as a parametric quadratic, and fit the curve $(u,v(u),w(u))$ using the MLS approximation. Finally, the 
MLS projection is computed by evaluating the curve at $u=0$. 

\begin{figure}[t!]
\centering
\epsfxsize=2.5in
\epsffile{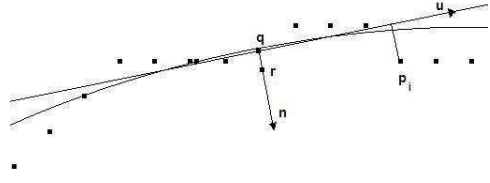} 
\caption{MLS Projection procedure for curves}
\label{mls}
\end{figure}

Though we assume that the surface under consideration does not have any sharp 
corners, we cannot make the same assumption about the boundary.  For instance 
a rectangular sheet that is segmented out has 4 sharp corners in the boundary 
though the interior is smooth. In order to preserve the 
sharp corners, the user is optionally permitted to specify the corners in the 
boundary so that each piece of the boundary is smoothed separately and the 
features preserved. 

\subsection{Filling Holes}
\label{sec:hole}
The MLS projection technique is used to fill holes in the input. Since a 
global parameterization of the data is not available, the filling procedure 
must be local to handle holes of arbitrary geometry. Hence we 
use the following method that is efficient and flexible \cite{TCH}. 

\begin{enumerate}
	\item For every pair of adjacent edges $b_{1}$ and $b_{2}$ in the 
boundary a new edge is introduced between the two edges, and hence a new 
triangle, if the angle between the $b_{1}$ and $b_{2}$ is less than $\phi$ . 
Further, this edge is introduced only if it does not intersect any other 
boundary edge locally.
	
	 Since new boundary edges are introduced in this process, multiple passes 
of this step are made until no adjacent edges make an angle less than 
$\phi$.
	\item For every edge $e$ in the new boundary
	\begin{enumerate}
		\item A local neighborhood of the mid-point of the edge and a local 
parameterization is found using the best fit plane using the MLS procedure. 
	  \item A new point along the perpendicular bisector of $e$ in the local 
parameterization is chosen, at a specified distance $d(e)$.
	\item A a new point on the surface is computed using a local MLS 
approximation.	
	\end{enumerate} 	
 	\item For every new point $p$, the closest edge in the current boundary 
is found. Let the end points of the edge be $e_{1}$ and $e_{2}$. In order to 
ensure that well behaved triangles are obtained, a check is made to see if 
introducing a new triangle $t$ with $p$, $e_{1}$ and $e_{2}$ crosses any 
other boundary edge in the local parametric domain. If so, the point $p$ is 
discarded. If not, a triangle $t$ is introduced.	
 	\item If the boundary size is greater than 3, the entire process is 
repeated again starting from step 1. If not, a new triangle is introduced 
with the three vertices left in the boundary and the process ends.
\end{enumerate}

\begin{figure}[h]
\begin{tabular}{c}

\subfigure[]{
\label{Hole Boundary}
\begin{minipage}[t]{0.23\textwidth}
\centerline{\epsfxsize=0.9\textwidth 
\epsfbox{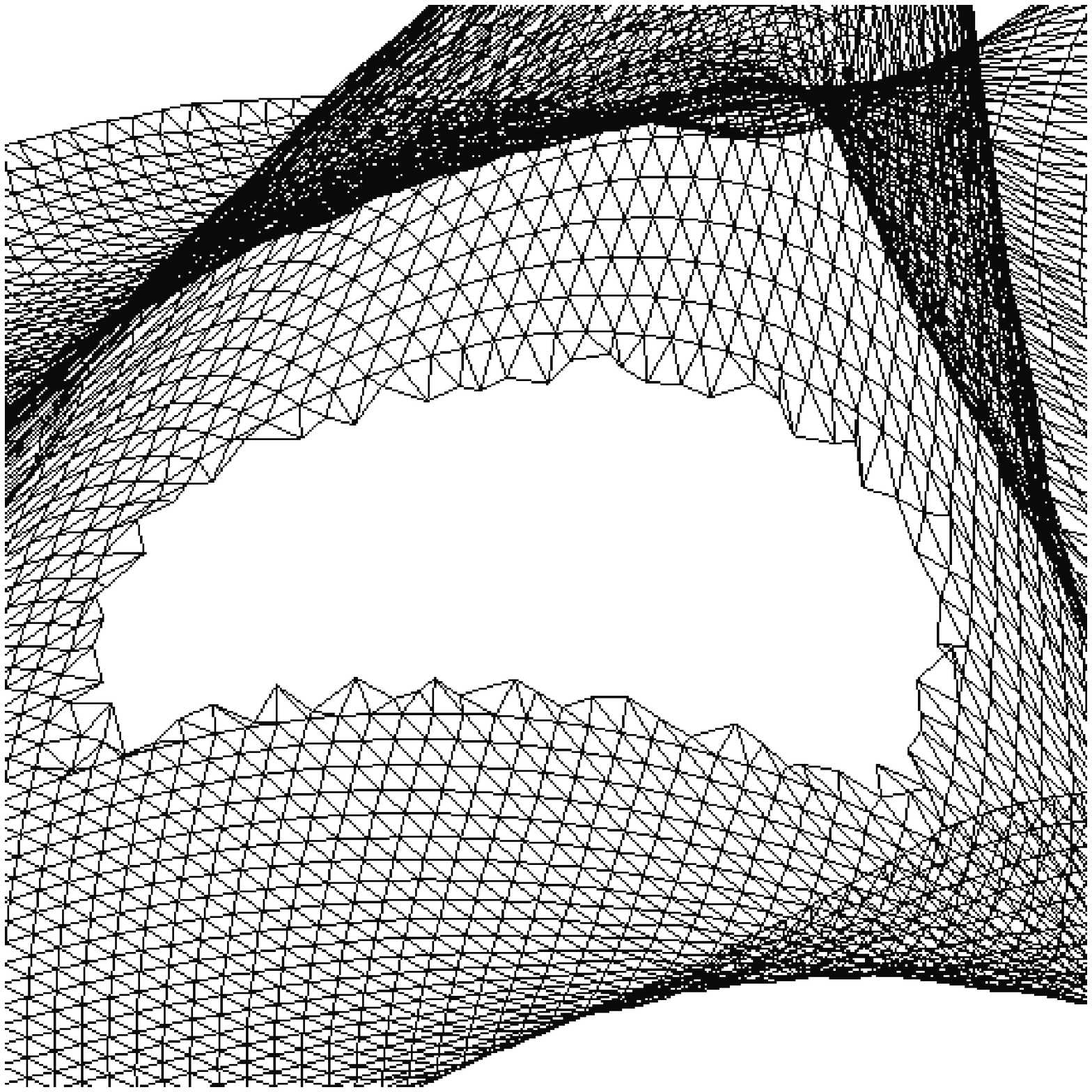}}
\end{minipage}}

\subfigure[]{
\label{hole_filled}

\begin{minipage}[t]{0.23\textwidth}
\centerline{\epsfxsize=0.9\textwidth 
\epsfbox{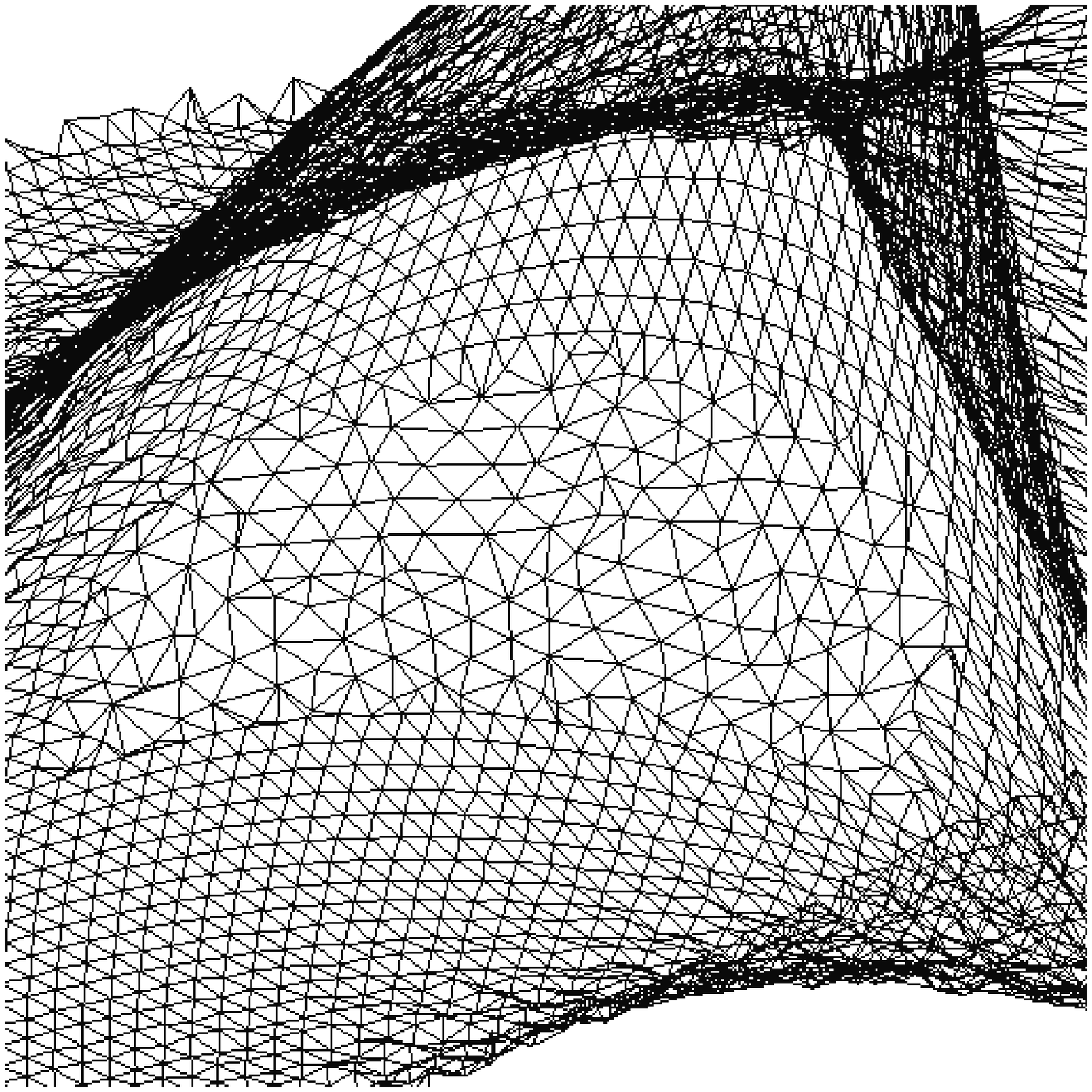}}
\end{minipage}}

\end{tabular}

\caption{A close up view of the hole, filled using our method}
\end{figure}
    
The choice of $d(e)$ and the value of $\phi$ play a significant role in 
determining the shape of the triangulation. If $\phi$ is too large, 
ill-shaped triangles result, that are elongated in one direction. We have 
obtained reasonable results for $\phi$ of about $5\pi/9$. Some alternatives 
for choosing $d(e)$ are to select a $d(e)$ so that the resultant triangle is 
equivalateral in the parametric domain with side $e$, where $e$ is the side 
under consideration, or to create an isosceles triangle in the parametric 
domain with two of the sides as the average edge length $a$. We choose the 
later method and choose $d(e)$ as $ \sqrt(4a^2-e^2)/2 $.
  
\subsection{Parameterization}
We use mean value coordinates \cite{F}, to parameterize data. This method is 
a discretization of harmonic maps that are based on the fact that harmonic 
maps satisfy the mean value theorem. The method proceeds by mapping the 
boundary of the mesh to a convex polygon and solving a linear system   of 
equations that express every point as a convex weighted average of its 
neighbors in the parametric domain where the weights are obtained by the 
application of mean value theorem for harmonic functions.


An important issue that arises while using convex combination maps for 
parameterization is fixing the parameter values of the boundary to a convex 
polygon. When dealing with rectangle shaped objects, in order to obtain an 
intuitive parameterization, we let the user specify the boundary points that 
are fixed to the boundary of a rectangle. In other cases, we map the boundary 
of the object directly to a square by chord length. 

\subsection{Knot Placement}
An important aspect of the fitting process is placing the adequate number of 
knots at the right locations. The knot placement strategy we use is to 
recursively subdivide the domain at the center of each region. The 
subdivision terminates when the data in the region can be fit locally with 
polynomial basis functions of degree 3, with an error no more than $\kappa$, where $\kappa$ is measured as the average of the square root of sum of squares of errors of each point, or until a specific recursion depth is reached.

\begin{figure}[!htp]

\centerline{\epsfxsize=0.4\textwidth 
\epsfbox{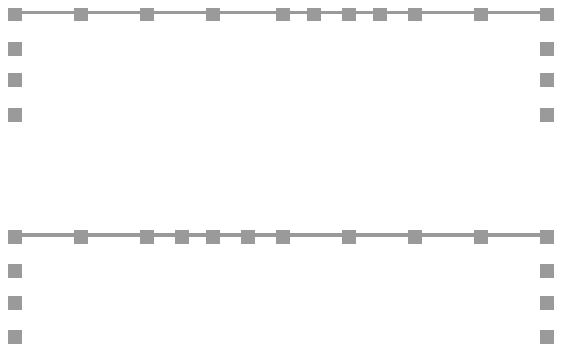}}
\caption{The knot vector obtained using hierarchical domain decomposition}
\label{knot-vector}

\centerline{\epsfxsize=0.3\textwidth 
\epsfbox{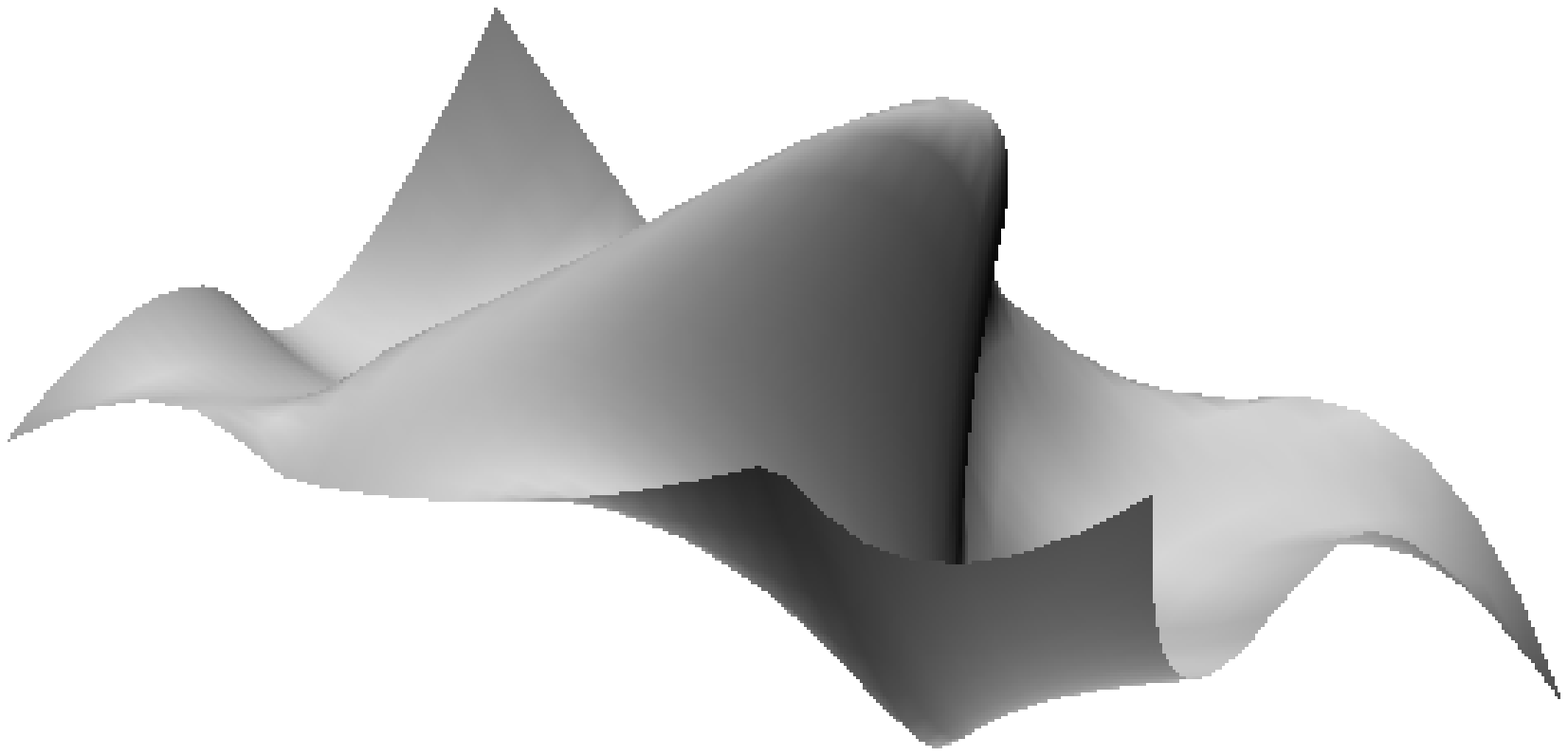}}
\caption{The surface under consideration}
\label{knot-vector}

\end{figure}

The knot vector finally contains the end points of all the patches along with additional knots added at the corners to make it open uniform. If there are $n_{p}$ patches in the $u$ direction, there are $n_{p}+1$ knots that are obtained from the end points of patches and hence there are $n_{p}+7$ knots in all when additional knots are added to make it open cubic.

\subsection{Fitting: Blending Local Fits}
The fitting step produces  a tensor product B-spline surface that  represents 
the shape of the input, given the parameterization over a rectangular domain. 
We propose a new method based on blending local fits to obtain a global fit. 

\begin{figure}[!h]
	\subfigure[]{
	\label{basis-whole}
	\epsfxsize=0.45\textwidth 					
	\epsfbox{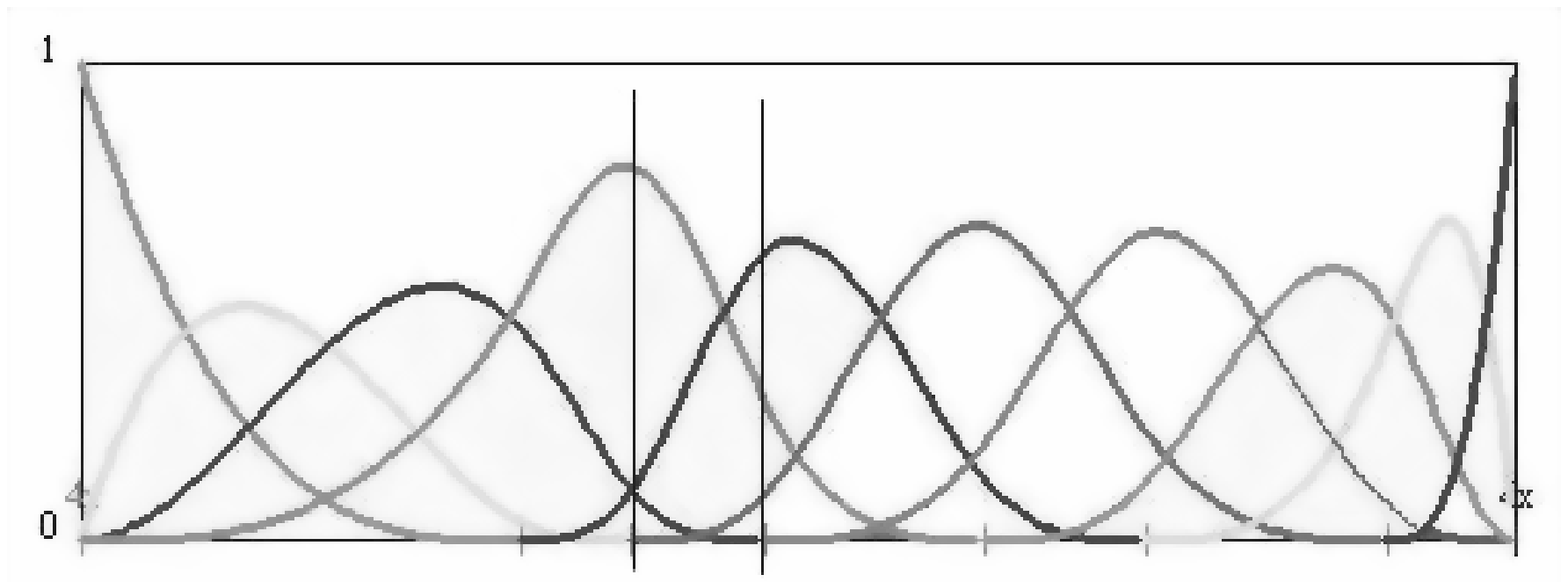}}
	\subfigure[]{
	\label{basis-isolated}
	\epsfxsize=0.425\textwidth		
	\epsfbox{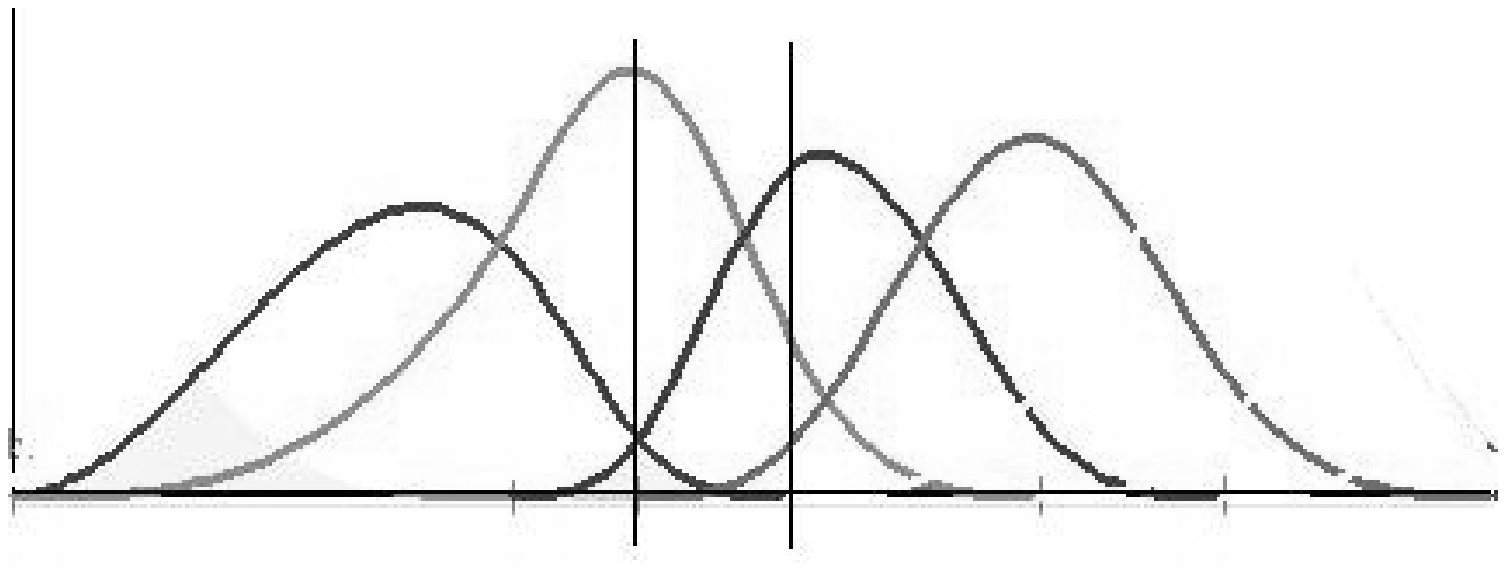}}

	\caption{The first figure shows the global basis functions corresponding to the knot vector	and the second figure shows the basis functions for a local fit in the particular interval }
		\label{basis}
\end{figure}



To illustrate our method, this section discusses the blending local fits method to fit B-spline curves. Suppose the hierarchical domain decomposition process leads to $n_{p}$ patches, the knot vector has a size of $n_{p}+7$, with $n_{p}+3$ basis functions. If the total number of data points is $N$, and we want to fit a B-spline curve $\gamma=(t,f(t))$
where $f(t)=\sum \limits_{j=0}^{n_{p}+2}{C_{j}\beta_{j,k}(t)}$, given data points $\{(t_{i},f_{i})\}_{i=0}^{N-1}$, global least squares fit minimizes 
$\sum \limits _{i=0}^{N-1} 
{(\sum \limits_{j=0}^{n_{p}+2}{C_{j}\beta_{j,k}(t_{i})}-f_{i})^{2}}$

Let the $n_{p}$ segments resulting from the domain decomposition process be $\{P_{i}\}_{i=0}^{n_{p}-1}$ and the mid-points of these patches be $\{M_{i}\}_{i=0}^{n_{p}-1}$ and the number of points in each patch be $\{N_{i}\}_{i=0}^{n_{p}-1}$. 

For each patch $P_{p}$, the coefficients of the local B-spline fit, $L_{0}^{p}$, $L_{1}^{p}$, $L_{2}^{p}$ and $L_{3}^{p}$ are found by applying the least squares criterion to the points in each patch by minimizing
$\sum \limits _{i \in N_{i}} 
{(\sum \limits_{j=0}^{3}{L_{j}^{p}\beta_{j+p,k}(t_{i})}-f_{i})^{2} w(t_{i}-M_{p})}$. where the function $w$ is shown in figure \ref{window}.

In other words, the four basis functions used for any patch are the four B-spline basis functions that are non-zero in that patch as shown in figure \ref{basis}.

The final curve is a blend of the local curve pieces that are fit independently, joined with $C^{2}$ continuity. 
We attempt constructing a global control mesh where control points of each patch coincide with the appropriate control points of the neighbouring patches by the process if averaging control points of neighbouring patches as described below.

If $\{G_{i}\}_{i=0}^{n_{p}+2}$ are the control points of the global mesh, \linebreak $G_{i} = {r_{0}L_{3}^{i-3}+r_{1}L_{2}^{i-2}+r_{2}L_{1}^{i-1}+r_{3}L_{0}^{i}}$ where $(r_{0}+r_{1}+r_{2}+r_{3})=1$.

The first and the last terms of the above expression have less significance compared to the terms at the center because the control point $G_{i}$ has lesser effect on patch $P_{i-3}$ compared to $G_{i-1}$ and $G_{i-2}$. Also, the weight given to points in the patch $P_{i}$ is small while doing the local moving least squares fit for patch $P_{i-1}$, as discussed later. So we use the simplified weights of $r_{0}=r_{3}=0$ and $r_{1}=r_{2}=1/2$.

Hence, in our scheme, the coefficients of the global control mesh are obtained as $G_{i}=\frac{L_{2}^{i-2}+L_{1}^{i-1}}{2}$ for $i$ from $2$ to $n_{p}+2$, $G_{0}=L_{0}^{0}$, $G_{1}=L_{1}^{0}$, $G_{n_{p}+1}=L_{2}^{n_{p}+1}$ and $G_{n_{p}+2}=L_{3}^{n_{p}+2}$. The coefficients of two adjacent segments ae shown in the figure \ref{coeffs}.

\begin{figure}[!h]
	\centerline{
	\epsfxsize=0.4\textwidth 					
	\epsfbox{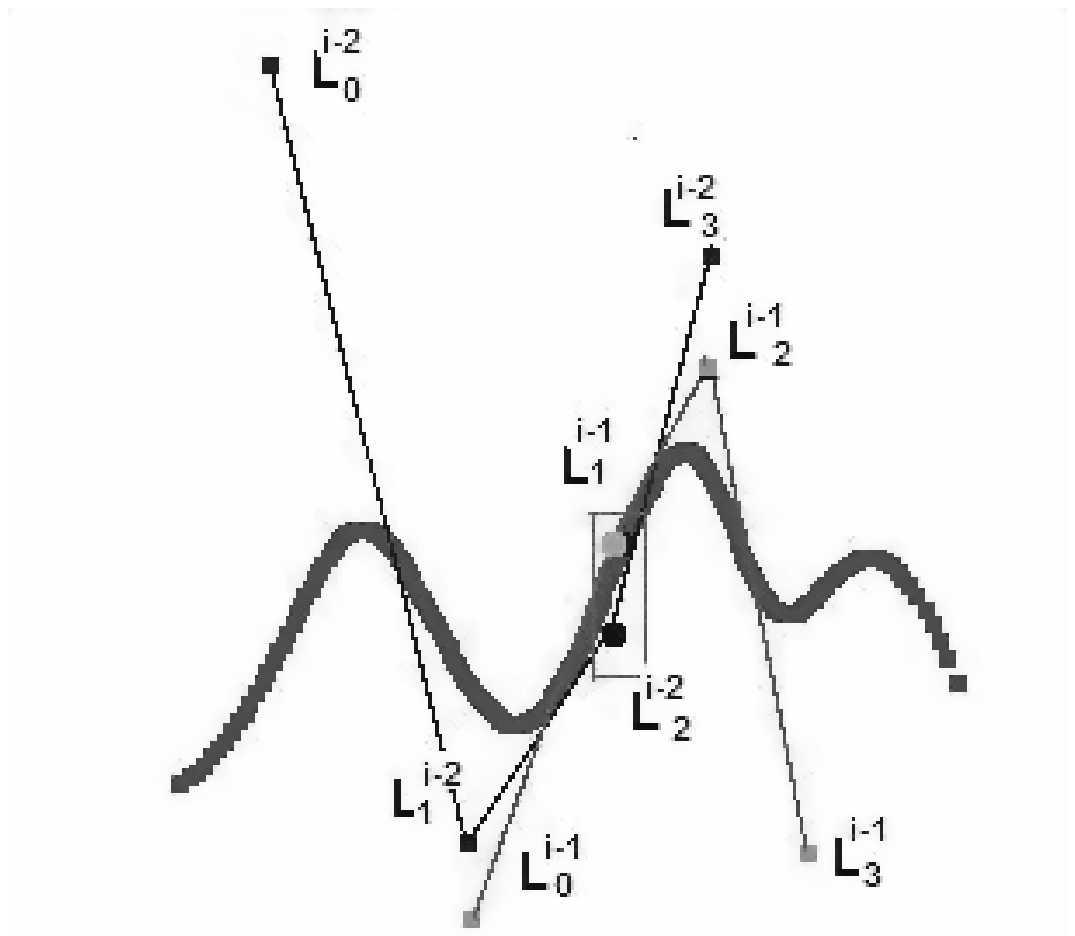}}
	
	\caption{A figure showing the control polygons of two local patches and the process of blending}
	\label{coeffs}
	\end{figure}


The local fit for a surface patch is done using local weighted least squares with respect to the 
mid-point of each patch in the parametric domain, in a way similar to the curve example. The basis functions used 
for the local fit are the tensor-product cubic B-spline basis functions that are non-zero in the 
interval under consideration. Since 16 basis functions are non-zero for a 
given knot interval, upon performing the local fit, 16 coefficients are 
obtained. These are the 16 control points of a local patch. In order to 
obtain the control points that constitute the global control mesh, the 
control points of four adjoining patches corresponding to the same location 
in the parametric domain, are blended. At the boundaries, the control points 
of each pair of adjoining patches are blended.

To fit each individual patch, we consider points from the patch under 
consideration and the adjoining patches to get a smooth blend. Considering 
more points from other patches, while giving a smoother blend, might fail to 
capture some features in the input. We attempt to control this effect by 
using an appropriate weighting function.

The choice of weighting function used in this process affects the result 
significantly. We use a function that gives more weight to the interval under 
consideration and partial weight to the neighboring patches so that the local 
surface blends smoothly into the neighboring patches. With a knot vector 
$\{t_{0} \mbox{,} t_{1} \mbox{...} t_{i} \mbox{,} t_{i+1} \mbox{, ...} 
t_{n}\}$, we use the the windowing function in the curve case.

$w(t) = \begin{cases}
 e^{\frac{{(t - t_{i})^2 }}{{(t_{i} - t_{i - 1})^2 /4}}{\rm}}& t<t_{i} \\ 
 1{\rm                     } & t(i) \le t \le t{i + 1} \\ 
 e^{\frac{{(t - t_{i + 1})^2 }}{{(t_{i + 1} - t_{i})^2 /4}}{\rm}} & t > t_{i 
+ 1} \\ 
 \end{cases} $

Hence, we give full weight to the interval under consideration and decrease 
the weight exponentially beyond the interval as shown in figure
\ref{window}. For surfaces, we construct two such windowing functions based on the knot vectors in the $u$ and $v$ directions and create a tensor product function.

This approach of taking a linear combination of control points to obtain the global mesh is not the optimal way of solving this problem and differs from the approach taken by the global least squares fit that tries to minimize the overall error of the fit rather than for each independent piece. A comparisn of both methods is given in the next section.

\begin{figure}
\begin{center}
\epsfxsize=2.5in
\epsfbox{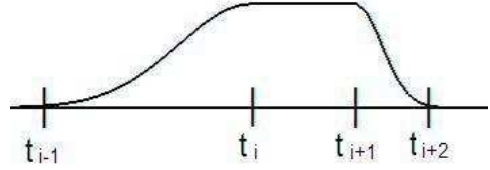}
\caption{The weighting function that windows the neighborhood.}
\label{window}
\end{center}
\end{figure}

\subsubsection{Analysis of Blending Least Squares Fit}
The preferred method for solving the linear least squares problem for data 
that is not well conditioned is the technique of singular value 
decomposition. This method takes a running time of the order $4\ell 
m^2-4m^3+O(\ell ^2)$ where there are $\ell$ data points and $m$ control 
points to solve for. The blending local fits method provides substantial 
speed up computationally by using a constant $m$ of 16 for each patch. Hence, if $a$ is the number of rows and $b$ the number of columns in the global control mesh, the blending local fits method solves $(ab-3a-3b+9)$ local systems and each local 
system has approximately $9\ell/ab$ points. In effect the Blending local fits method reduces the runtime complexity of B-spline surface fitting from 
$O(\ell m^2)$ to $O(\ell)$. Hence the size of the control mesh does not have much effect on the complexity of the fitting algorithm and considerable 
speedup can be realized for large control meshes.

\begin{figure}[!h]
\begin{center}

\subfigure[]{
\label{local-error}
\centerline{\epsfxsize=0.4\textwidth 
\epsfbox{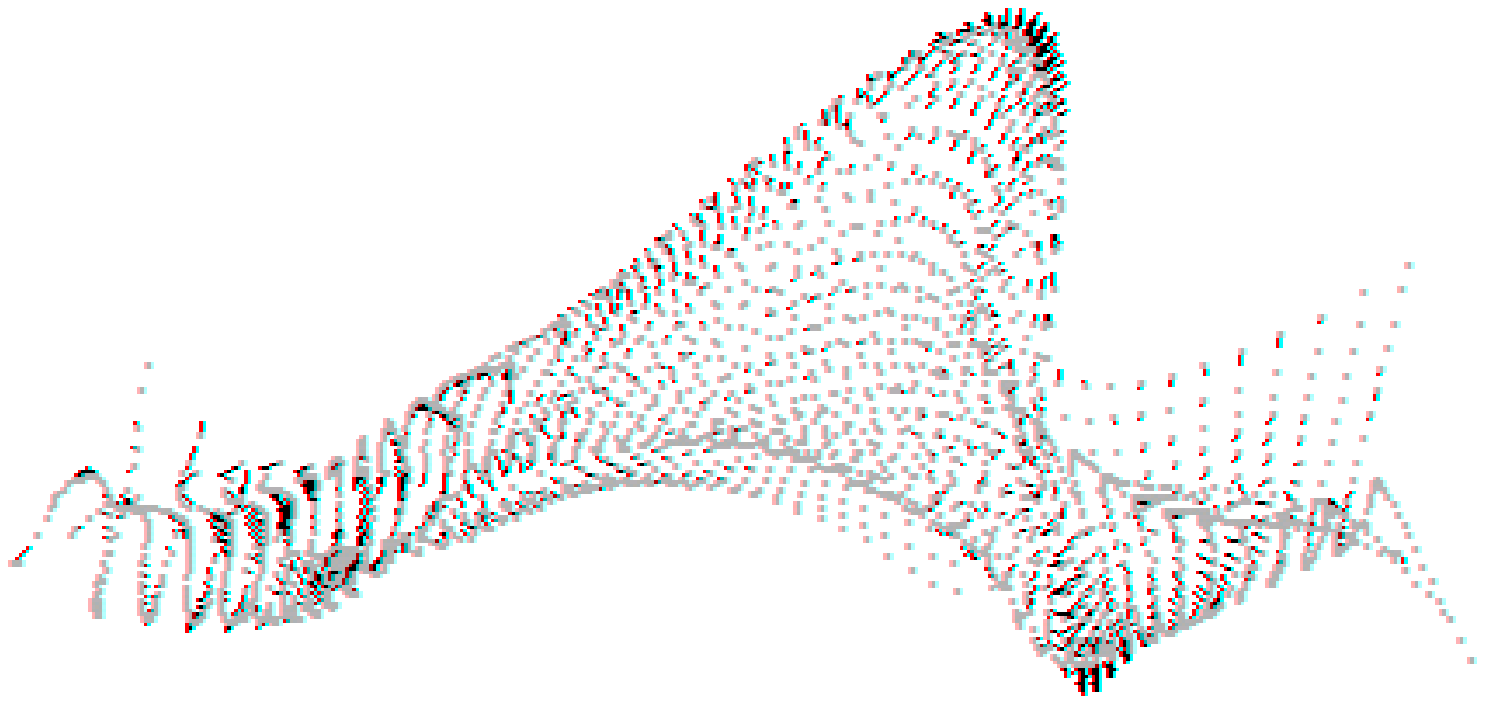}}}

\subfigure[]{
\label{global-error}
\centerline{\epsfxsize=0.4\textwidth	
\epsfbox{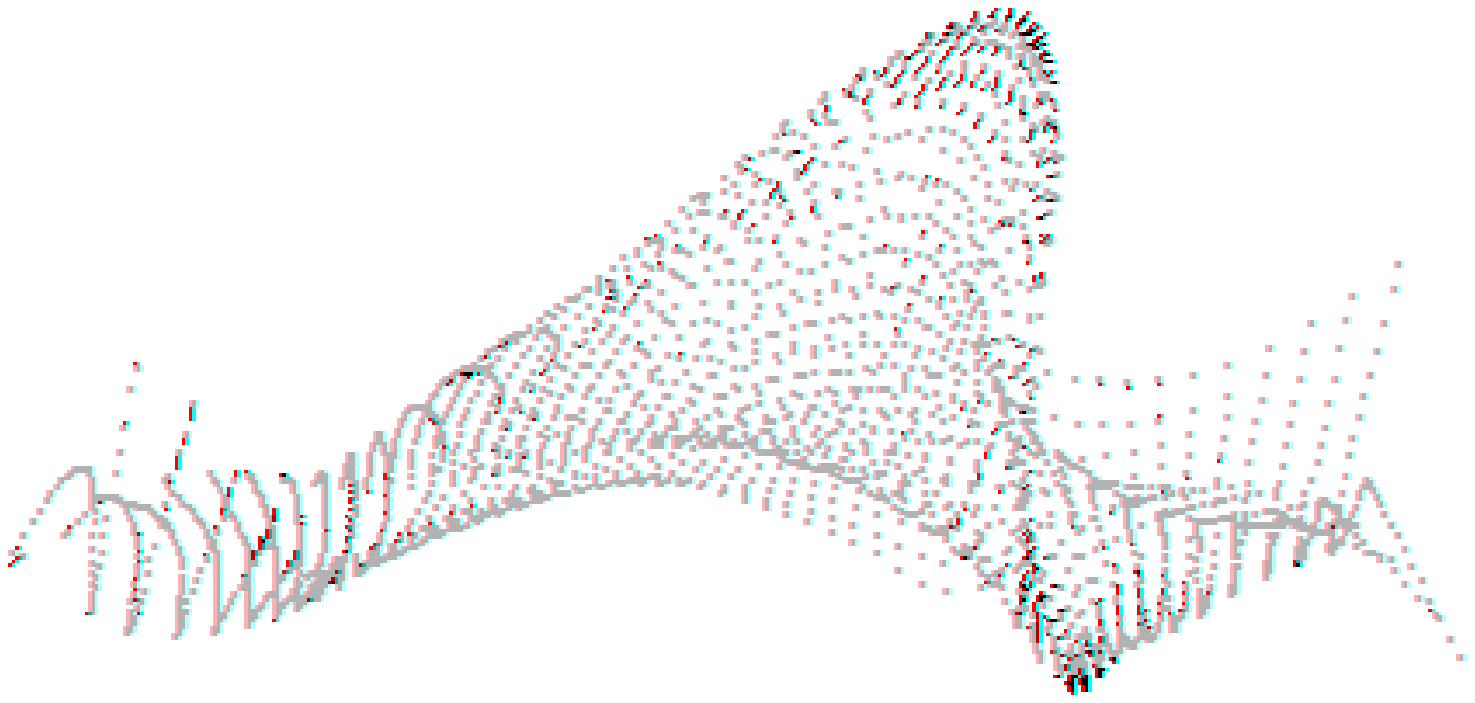}}}

\caption{Lines indicating the distance between points on the fitted surface 
and the actual points for the same parameter value using the blending local 
fits method and the global least squares fit method.}

\label{error}
\end{center}
\end{figure}

The quality of fit obtained using the blending local fits method largely 
depends on size of the neighborhood chosen and the weighting function used. 
Figures \ref{local-error} and \ref{global-error} show the error obtained when approximating the data using the blending local fits method and the global least squares fit method respectively, with a 13x13 control grid for 2500 points. We have taken a simpler version of our original input that we use later in order to perform the comparisn in a reasonable amount of time, without the requirement of optimizing the fitting code to exploit the sparseness of the linear system. We obtain a fitting error of 0.098 per point while the global least squares fit achieves a 0.047 fitting error per point using the same knot vector. This is small compared to the feature size as shown in figure \ref{error}. Though a global least squares fit leads to a lower error than our fitting method, substantial computation time can be saved by using our fitting method.

\section{Results}

\begin{figure}[!h]

\centerline{\epsfxsize=0.38\textwidth 
\epsfbox{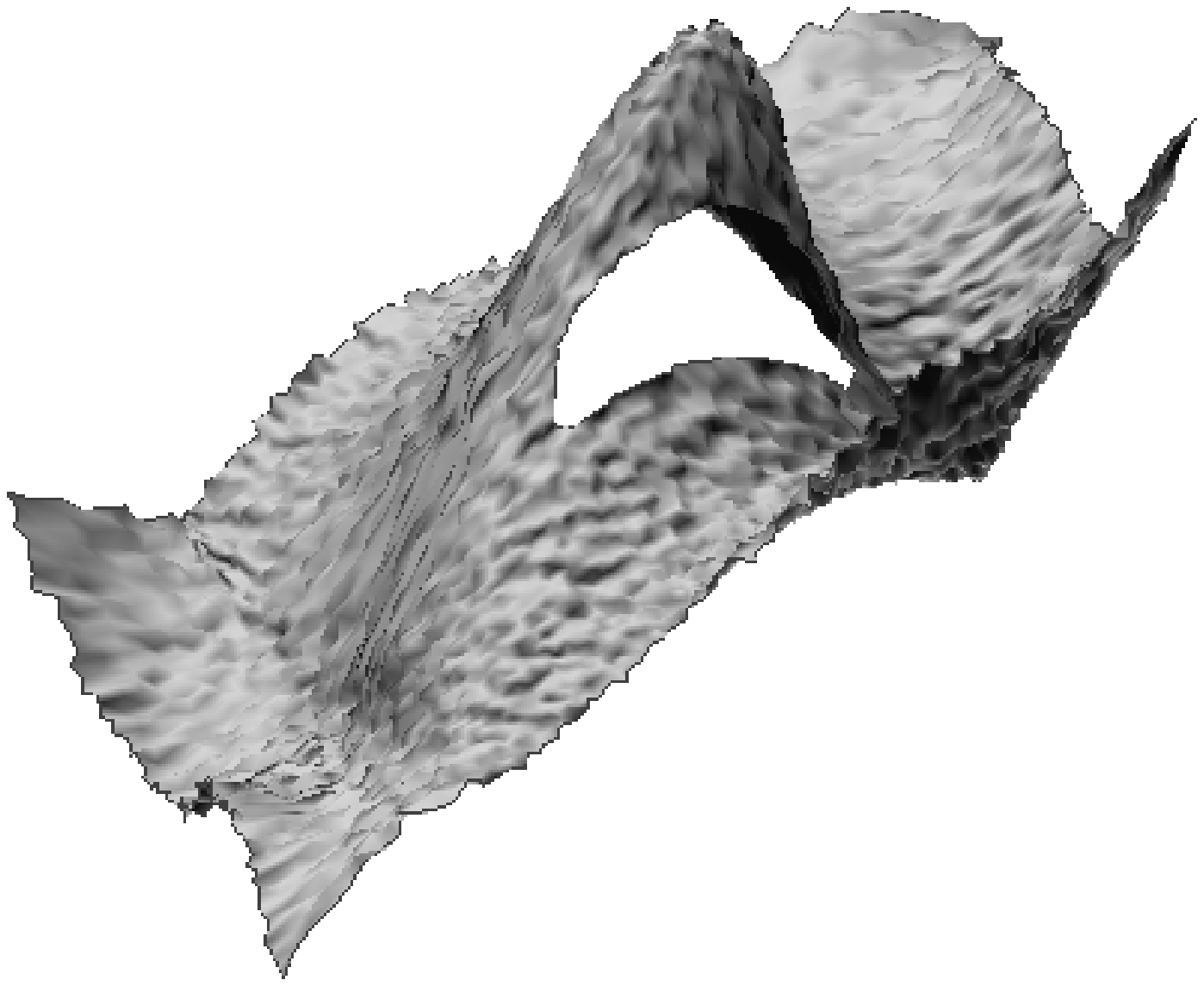}}
\caption{Noisy and incomplete data}
\label{noisy}

\centerline{\epsfxsize=0.38\textwidth
\epsfbox{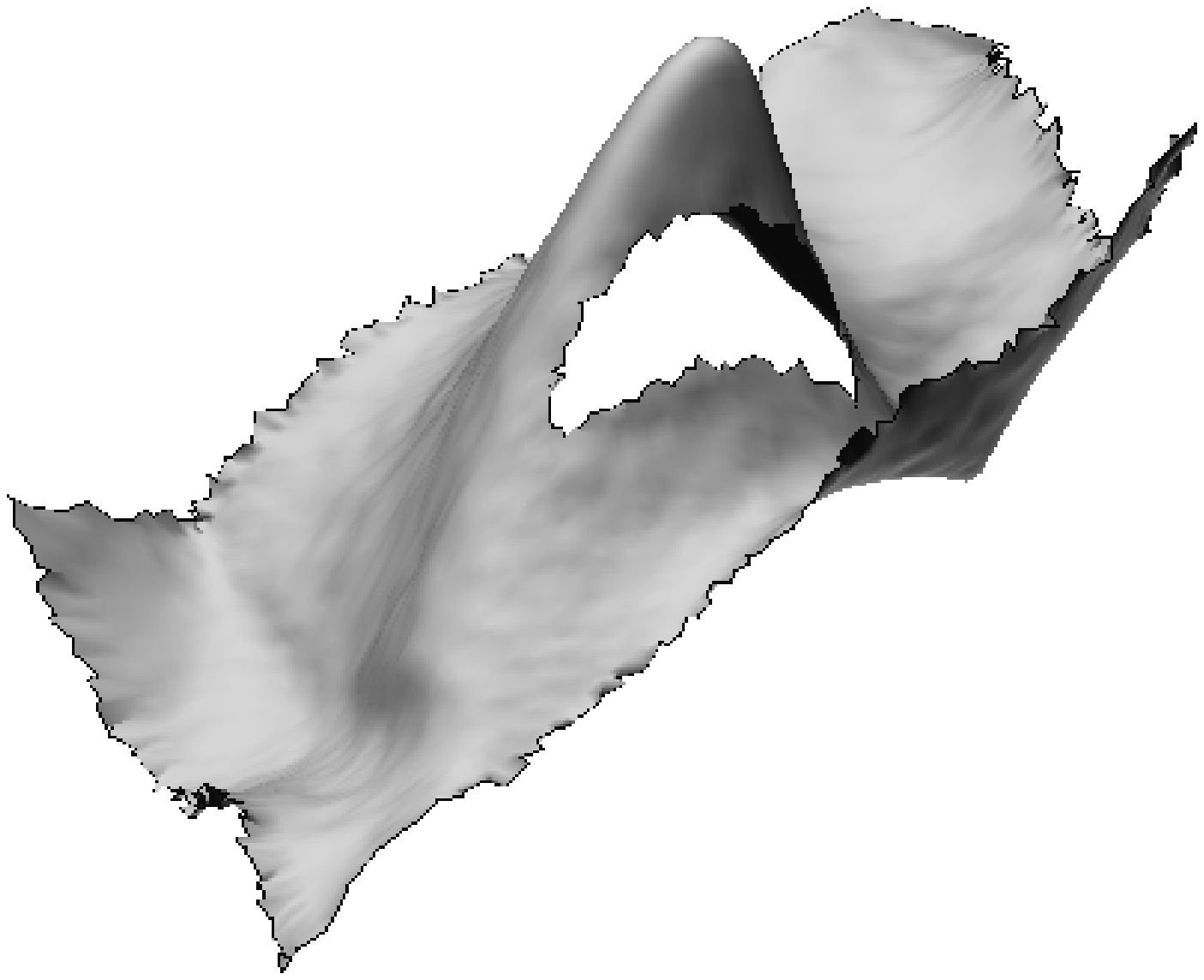}}
\caption{Smoothed data, without the boundaries smoothed}
\label{smoothed}

\centerline{\epsfxsize=0.38\textwidth
\epsfbox{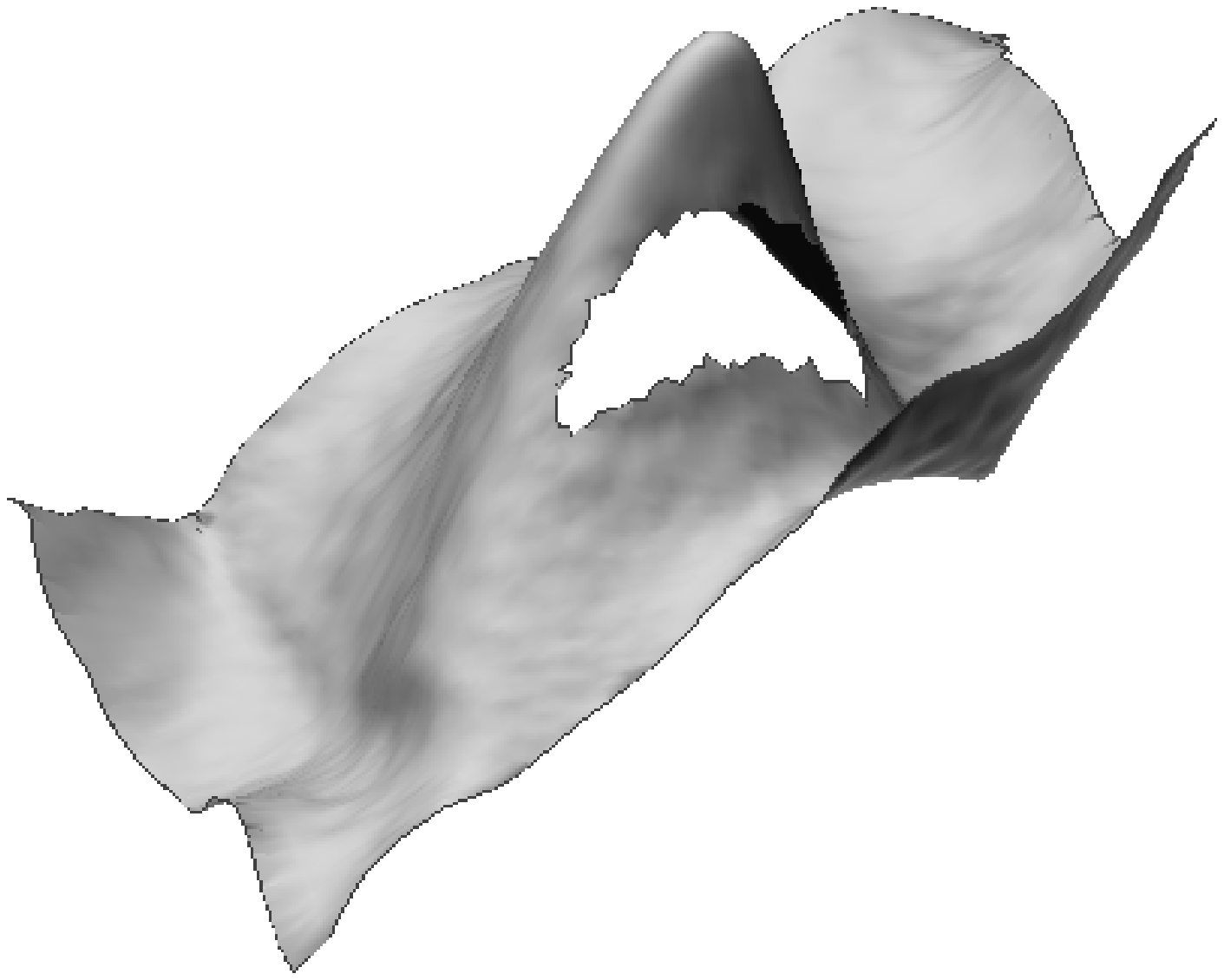} }
\caption{The smoothed surface}
\label{smooth-all}

\end{figure}

\begin{figure}[h!]

\centerline{\epsfxsize=0.38\textwidth
\epsfbox{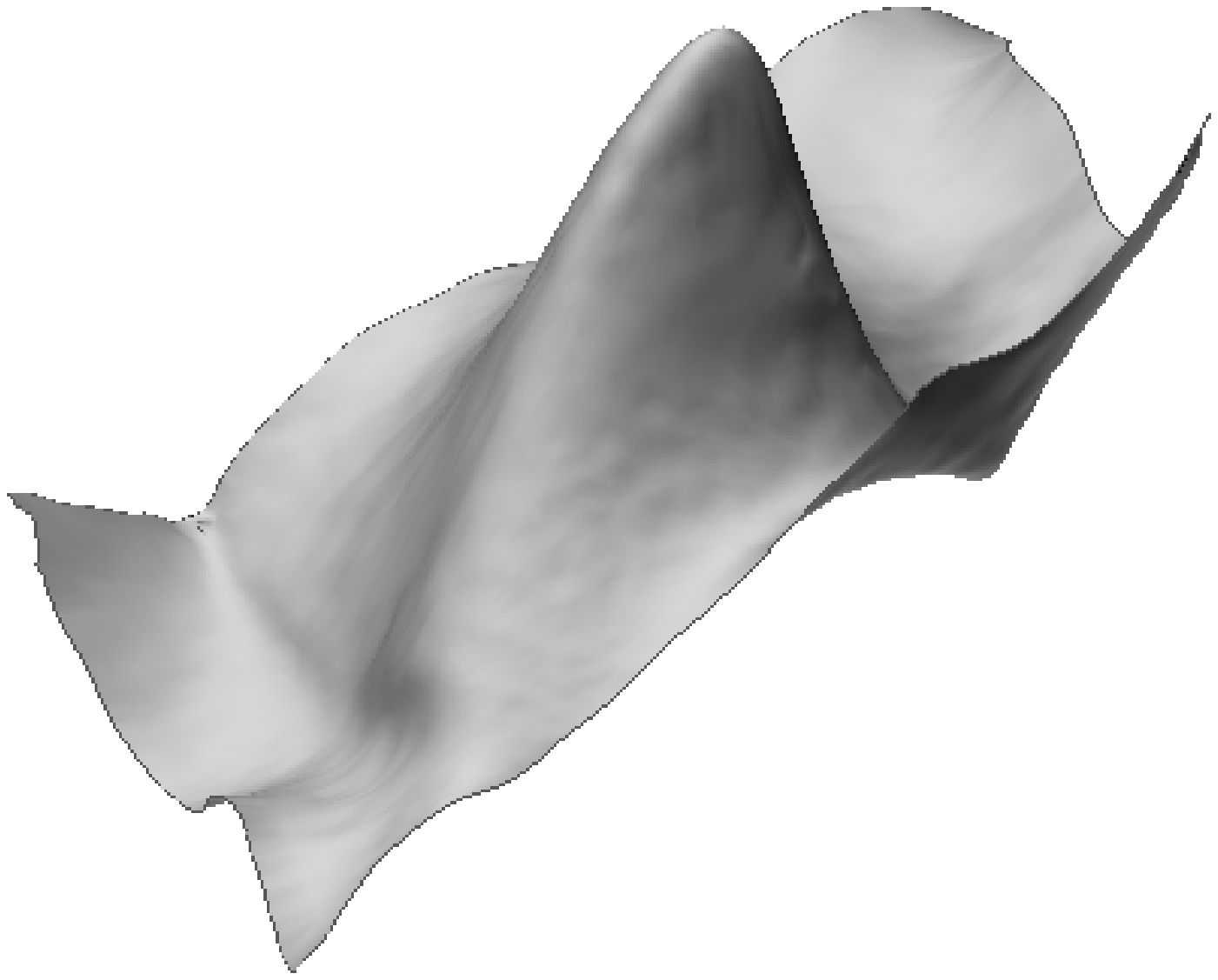} }
\caption{Hole filled}
\label{filled-full}

\centerline{\epsfxsize=0.38\textwidth
\epsfbox{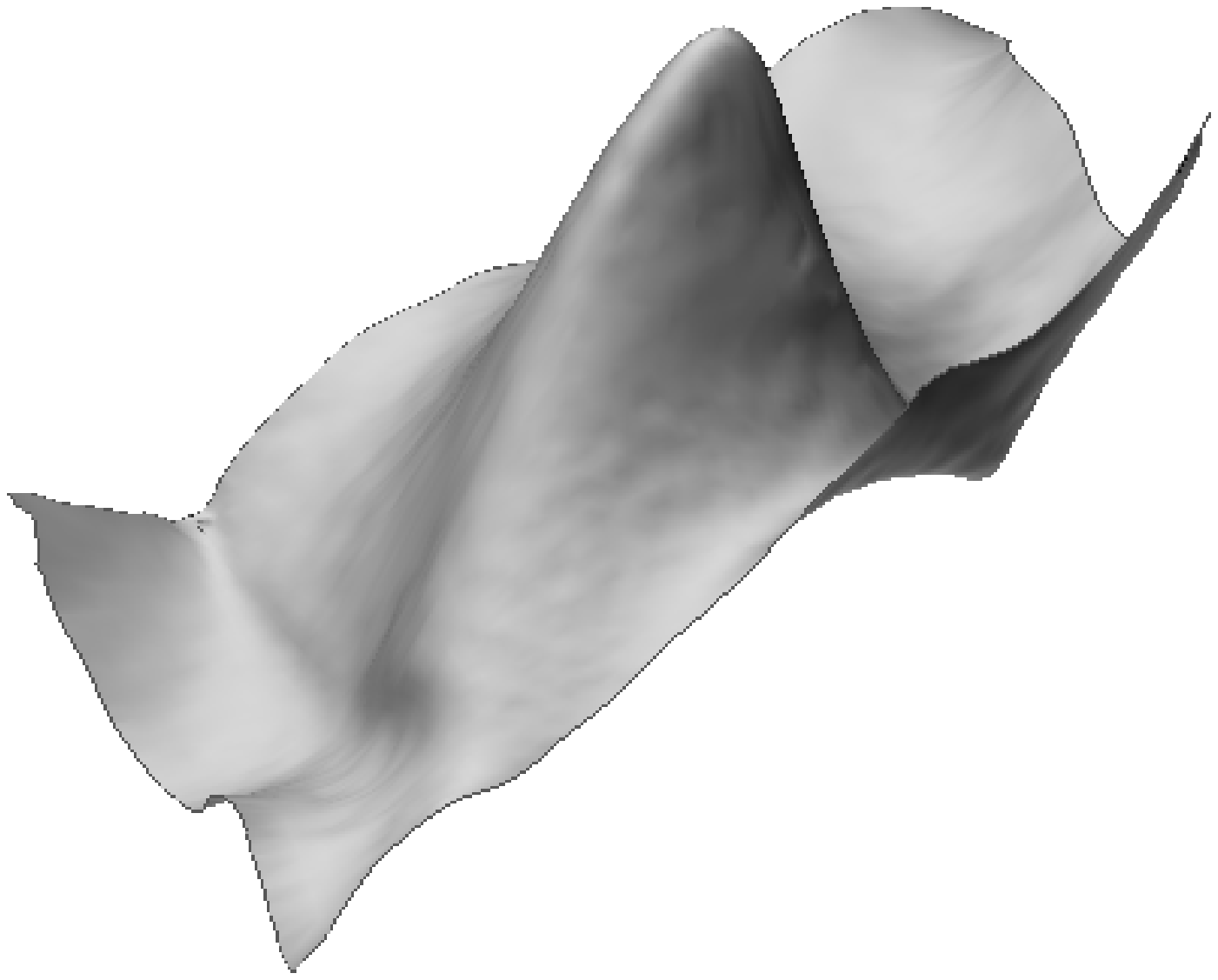} }
\caption{B-spline surface fit}
\label{fit}

\centerline{\epsfxsize=0.3\textwidth
\epsfbox{knot_vector.eps}}
\caption{Hole filled}
\label{knot-vector1}

\end{figure}

This section demonstrates the entire framework on the noisy data shown in 
figure \ref{noisy} with 8100 points. Figure \ref{smoothed} shows the noise at the 
boundary is removed by smoothing the boundary as shown in figure 
\ref{smooth-all}. 
The hole in the input is then filled using the hole filling algorithm shown 
in section \ref{sec:hole}. The final fit, with a $13\times13$ control mesh, 
as shown in figure \ref{fit} is obtained using the blending local fits 
method. The knot vector obtained using the adaptive domain decomposition 
technique is shown in figure \ref{knot-vector1}.

\section{Conclusion}
This work outlines a framework to convert point cloud data with 
moderate complexity with associated triangulation information into tensor 
product B-spline surfaces or to fit a patch of the original segmented data 
set. This includes the multi-step approach of smoothing, hole filling, 
parameterization and finally fitting the surface. Though several patch-based 
methods and softwares already exist to solve the reverse engineering problem, 
this method aims to deal with point clouds that have problems like noise, 
holes in the geometry and missing data. Also this method aims to capture more 
detail in a single patch by deciding where to place the knots using a 
hierarchical subdivision of the domain and uses a combination of local 
weighted least squares approximations to find the control points of the 
tensor product surface as a whole. 
{\small{

}}}
\end{document}